\documentclass[11pt,preprint]{aastex}
\usepackage{graphicx}
\usepackage{graphics}
\usepackage{color}
\usepackage{epsfig}
\usepackage{epsf}

\def\spitzer{{\sl Spitzer}}
\def\hst{{\sl HST}}
\def\chandra{{\sl Chandra}}

\def\xs{NGC~5866}
\shortauthors{}
\shorttitle{}
\begin{document}

\title{Dynamic S0 Galaxies: a Case Study of NGC~5866}

\author{Jiang-Tao Li\altaffilmark{1,2},
Q. Daniel Wang\altaffilmark{2}, Zhiyuan Li\altaffilmark{2,3}, and
Yang Chen\altaffilmark{1}} \altaffiltext{1}{Department of Astronomy,
Nanjing University, 22 Hankou Road, Nanjing 210093, P. R. China}
\altaffiltext{2}{Department of Astronomy, University of
Massachusetts, 710 North Pleasant Street, Amherst, MA 01003, U.S.A.}
\altaffiltext{3}{Harvard-Smithsonian Center for Astrophysics, 60
Garden Street, Cambridge, MA 02138}

\begin{abstract}
S0 galaxies are often thought to be passively evolved from spirals
after star formation is quenched. To explore what is actually
occurring in such galaxies, we present a multi-wavelength case study
of NGC~5866 --- a nearby edge-on S0 galaxy in a relatively isolated
environment. This study shows strong evidence for dynamic activities
in the interstellar medium, which are most likely driven by
supernova explosions in the galactic disk and bulge. Understanding
these activities can have strong implications for studying the
evolution of such galaxies. We utilize {\sl Chandra}, {\sl HST}, and
{\sl Spitzer} data as well as ground-based observations to
characterize the content, structure, and physical state of the
medium and its interplay with the stellar component in NGC~5866. A
cold gas disk is detected with an exponential scale height of
$\sim10^2\rm~pc$. Numerous distinct off-disk dusty spurs are also
clearly present: prominent ones can extend as far as
$\sim3\times10^2\rm~pc$ from the galactic plane and are probably
produced by {\sl individual} SNe, whereas faint filaments can have
$\sim\rm kpc$ scale and are likely produced by SNe collectively in
the disk/bulge. We also detect substantial amounts of diffuse
H$\alpha$- and P$\alpha$-emitting gas with a comparable scale height
as the cold gas. We find that the heating of the dust and warm
ionized gas cannot be explained by the radiation from evolved stars
alone, strongly indicating the presence of young stars in the
galactic disk, though at a slow formation rate of $\sim 0.05
{\rm~M_\odot~yr^{-1}}$. We further reveal the presence of diffuse
X-ray-emitting hot gas, which extends as far as 3.5~kpc away from
the galactic plane and can be heated easily by Type Ia SNe in the
bulge.
% and shows a similar distribution as the old stars.
However, the mean temperature of this gas is $\sim 0.2\rm~keV$,
substantially lower than what might be expected from the mass-loss
of evolved stars and Type Ia SNe heating alone in the galaxy,
indicating that the mass loading from the cool gas is important. The
total masses of the cold, warm ionized, and hot gases are
$\sim5\times10^8\rm~M_\odot$, $4\times10^4\rm~M_\odot$ and
$3\times10^7\rm~M_\odot$, respectively. The relative richness of the
gases, apparently undergoing circulations between the disk and halo
of the galaxy, is perhaps a result of its relative isolation.
\end{abstract}

\keywords{galaxies: general-galaxies: individual
(NGC~5866)-galaxies: S0}

\section{Introduction}\label{sec:introduction}

Living at the intersection of spirals and ellipticals on the Hubble
tuning fork, lenticular (or S0) galaxies are often thought to be
remnants of spirals after star formation (SF) has ceased (or has
slowed down substantially). This transition in the SF rate (SFR)
must be directly related to the change of cold gas content and/or
density in the disk of such a galaxy. The change may be caused by
such processes as (e.g., Martig et al. 2009 and references therein):
ram-pressure stripping (of cold gas), termination or strangulation
(removing surrounding hot gas, hence a source of new cold gas),
morphological quenching (heating of a stellar disk or even
transforming to a spheroid), and exhaustion (consuming gas via SF).
While the ram-pressure stripping of the cold or hot gas is expected
to work primarily in a dense environment (in galaxy clusters and
possibly in rich groups), the morphological quenching via
galaxy-galaxy mergers should be most effective in groups. The
exhaustion, an internal process, can occur in both rich and poor
environments. For example, this process may occur for a relatively
isolated galaxy when its halo mass $M_h$ grows to be $\gtrsim
10^{11}\rm~M_\odot$ and its gas accretion mode turns from cold to
hot (e.g., Kere\v{s} et al. 2005; Kere\v{s} \& Hernquist 2009 and
references therein). In general, more than one of these processes
may be involved in the transition of a galaxy. Their relative
importance cannot be determined easily; great uncertainty remains in
our understanding of the processes, especially in terms of their
interplay with various other galactic activities such as the
feedback from stars and AGNs. While observational studies of galaxy
transformation have been concentrated on stellar properties,
theoretical efforts (or computational simulations) are primarily on
gaseous components on galaxy scales. This gap must be bridged to
make significant progress in the field.

Here we present a case study of the nearby edge-on S0 galaxy
NGC~5866 (also referred as M~102 or "Spindle" galaxy; see
Table~\ref{table:basicpara} for the salient parameters). Our goal is
to understand the potential role of the stellar energy and mass
feedback in regulating the evolution of various gaseous components
in this relatively isolated galaxy. NGC~5866 belongs to the poor
group LGG~396 with a velocity dispersion of $\sim 74
{\rm~km~s^{-1}}$ (e.g., Mulchaey et al. 2003). Two gas-rich
companions, NGC~5907 (Sc) and NGC~5879 (Sbc), are known to be
located several hundred kpc away; their interactions with NGC~5866
should be minimal. So we can focus on internal activities of the
galaxy. As will be discussed later, this isolated environment in the
``field'' may explain the relatively richness of molecular gas in
the galaxy  (Table \ref{table:basicpara}; Welch \& Sage 2003; Sage
\& Welch 2006) and its slow evolution with ongoing low-level SF many
gigayears after a major merger. There is also no evidence for strong
nuclear activity in the galaxy (Terashima \& Wilson 2003); the
nucleus is classified in optical to be an intermediate object
between a weak \ion{H}{2} region and a LINER (Ho, Filippenko \&
Sargent 1997). Therefore, we can examine the effect of the stellar
feedback with little confusion from other potential sources of
mechanical energy inputs. Furthermore, the nearly complete edge-on
inclination of the galaxy is optimal for the study of the galactic
disk/halo interaction. Indeed, archival {\sl HST} optical images
(first studied by Cantiello, Blakeslee \& Raimondo 2007 for globular
clusters) show numerous off-disk cold gas spurs, in addition to a
well-defined dust lane. To our knowledge, no characterization of
such spurs has ever been made for any S0 galaxies, let alone their
nature and role in the galactic gaseous evolution. A previous X-ray
study based on {\sl ROSAT} observation revealed that NGC~5866 is an
X-ray faint galaxy and that the X-ray emission is dominated by
stellar sources (Pellegrini 1994). There was also a study based on a
snap-shot (2~ks) observation made with {\sl Chandra} to see if an
AGN is present in the galaxy (Terashima \& Wilson 2003); only an
upper limit of $\sim4\times10^{38}\rm~ergs~s^{-1}$ was obtained,
assuming the distance of the galaxy to be $\sim15.3\rm~Mpc$
(Pellegrini 2005). Our study is based on a later 34~ks {\sl Chandra}
observation (previously studied by David et al. 2006 in a sample of
early-type galaxies), which allows us to resolve out much of the
discrete source contribution and to examine the potentially diffuse
X-ray emission from the galaxy. In short, with various available
tracers (optical absorption as well as H$\alpha$/P$\alpha$, infrared
(IR) and soft X-ray emissions), we can now characterize the cold
molecular/atomic, warm ionized, dusty and hot phases of the
interstellar medium (ISM) and explore their relationships with the
stellar feedback in the galaxy.

The organization of the paper is as follows. In \S\ref{sec:data}, we
briefly describe various data sets used for our examinations (mainly
\chandra, \hst, and \spitzer). Spatial and spectral analysis of
these multi-wavelength data is presented in \S\ref{sec:analysis}. We
discuss our results and their implications for understanding the
energetics and dynamics of the gas phases in \S\ref{sec:discussion}.
We summarize the results and conclusions  in \S\ref{sec:summary}.

\section{Observation and Data Reduction}\label{sec:data}

\begin{figure}[!h]
\begin{center}
\epsfig{figure=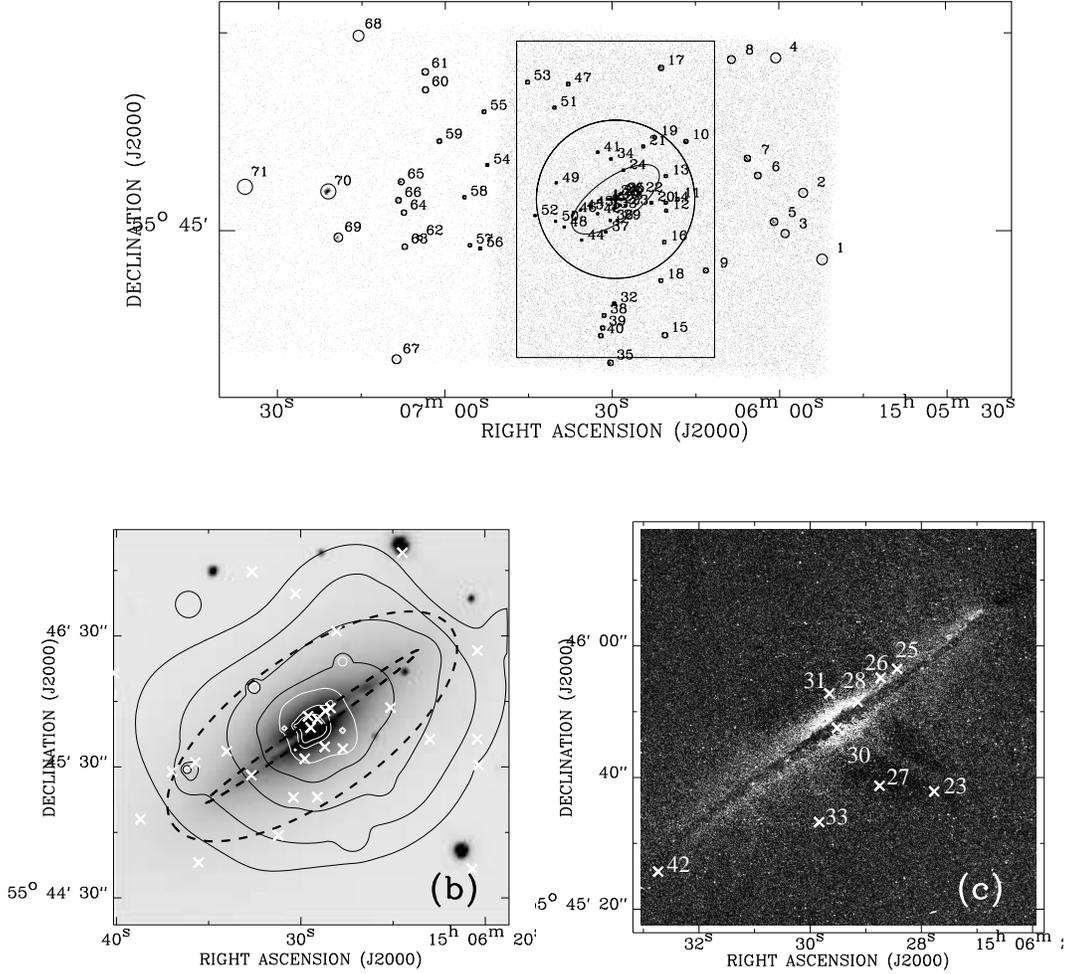,width=0.85\textwidth,angle=0, clip=}
\caption{(a) \chandra/ACIS-S image of NGC~5866 in the 0.3-7 keV
band. Detected sources are enclosed with circles of about 96\% PSF
(point spread function) EER and labeled with the numbers as in
Table~\ref{table:acis_source_list}. The plus sign marks the center
of the galaxy. While the ellipse marks the bulge region used for the
spectral analysis, the field between the large circle and box is for
the local sky background. (b) The central $3^\prime\times3^\prime$
($9.7\times9.7\rm~kpc$) of NGC~5866. Intensity contours of the
source-removed emission in the ACIS-S 0.5-1.5 keV band overlaid on
the SDSS R-band image of NGC~5866; the contours are at $(2, 3, 5,
10, 20, 55, 75)\times10^{-3}\rm~cts~s^{-1}~arcmin^{-2}$. The removed
point sources are marked by crosses. The large dashed ellipse
outlines the bulge region as in (a), while the small ellipse marks
the region used to calculate the galactic disk mass. (c) H$\alpha$
emission in the central region (the scale is the same as that in
(b)) as described in \S\ref{sec:data} with X-ray sources marked as
white crosses.}\label{fig:pointsrc}
\end{center}
\end{figure}

The {\sl Chandra} observation of NGC~5866 was taken on Nov. 14,
2002. We reprocess the archived data for our study. After the
removal of background flares, a total effective exposure of 24.3 ks
remains. A "stowed background", normalized by the count rate in the
10-12 keV range, is used to account for the non-X-ray event
contribution. Our study is based on the data from the ACIS-S3 chip,
although part of the S2 chip data is also used for our source
detection, which is performed in the broad (B, $0.3-7\rm~keV$), soft
(S, $0.3-1.5\rm~keV$) and hard (H, $1.5-7\rm~keV$) bands (Fig.
\ref{fig:pointsrc}), following the procedure detailed in Wang
(2004). To study the diffuse X-ray emission, we remove detected
discrete sources from the data. Circular regions are  excluded
within twice the 90 per cent energy enclosed radius (EER) around
each source of a count rate $(CR)\lesssim 0.01\rm~cts~s^{-1}$. For
brighter sources, the source removal radius is further multiplied by
a factor of $1+log(CR/0.01)$. Generally about 96 per cent of the
source counts are excluded in such a removal. For our spectral
analysis of the diffuse X-ray emission, we further remove a local
sky background. This background spectrum is first extracted from the
source-removed data in a region surrounding the galaxy
(Fig.~\ref{fig:pointsrc}). The spectrum, with the stowed background
subtracted, is modeled with a thermal plasma model (MEKAL in XSPEC)
and a power-law (with a photon index of 1.4, Lumb et al. 2002),
representing both the Galactic and extragalactic background
contributions. This local sky background model, normalized for the
relative area coverage, is to be subtracted from on-source spectral
data.

We use {\sl HST} ACS (pixel size$\sim0\farcs05$) and NICMOS (pixel
size$\sim0\farcs2$) observations in both broad and narrow bands. An
H$\alpha$ image is constructed from the ACS on-line narrow band
(F658N) and off-line broad band (F625W) observations. The off-line
image, used for the continuum subtraction, is normalized by a factor
of 0.09 to account for the bandwidth difference, which is determined
by the average flux ratio of several bright foreground stars in the
field (thought to have no H$\alpha$ line emission). We further
construct a P$\alpha$ image from NICMOS data. The F190N continuum
observation is directly subtracted from the on-line F187N
observation because the two bands have approximately the same
bandwidth; there is no bright foreground stars found in the small
field of view in the NICMOS observations.

We further use archival {\sl Spitzer} IRAC 3.6~$\rm \mu m$, 8~$\rm
\mu m$ and MIPS 24~$\rm \mu m$ images as well as 2MASS near-IR data.
The 24~$\rm \mu m$ image is used to trace the dust emission, the
stellar-light-subtracted 8~$\rm \mu m$ image is used to trace the
PAH emission, while the 3.6~$\rm \mu m$ image and the 2MASS images
are used to trace the emission from old stars. We make an aperture
correction for the {\sl Spitzer} images, following Dale et al.
(2007). To subtract the old stellar contribution from the 8~$\rm \mu
m$ emission, we adopt a ratio of 4.23 between the stellar light in
3.6~$\rm \mu m$ and 8~$\rm \mu m$, since 3.6~$\rm \mu m$ emission in
galaxies are dominated by old stellar light (Hunter et al. 2006).
The pixel size of IRAC ($1\farcs2$) is smaller than that of MIPS
24~$\rm \mu m$ image ($2\farcs55$), so we rebin the 24~$\rm \mu m$
image to have the same pixel scale as the 3.6~$\rm \mu m$ image (but
note that the resolution of MIPS at 24~$\rm \mu m$ is only
$\sim5\farcs9$), and plot them in Fig.~\ref{fig:tricolorimgs}c. The
2MASS near-IR image (with a spatial resolution of
$\sim1^{\prime\prime}$) is also dominated by old stellar light in
galaxies, and the J-K color, as detailed in \S\ref{subsec:SF}, is
adopted as a tracer of the intrinsic extinction near the disk.

\section{Analysis and Results}\label{sec:analysis}

\begin{figure}[!htb]
\begin{center}
\centerline{
    \epsfig{figure=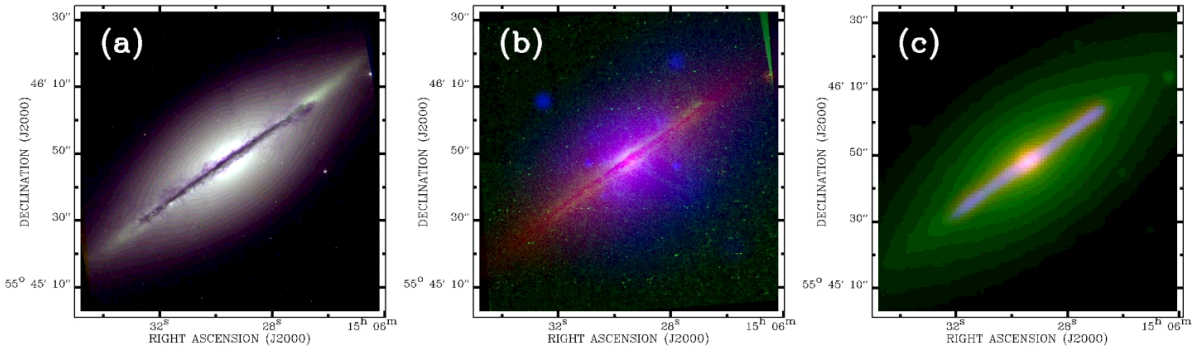,width=1.0\textwidth,angle=0, clip=}
} \caption{Tri-color images of the central $1\farcm5\times1\farcm5$
($4.9\times4.9\rm~kpc$) of NGC~5866: (a) {\sl HST} ACS F555W ({\sl
Red}), F435W ({\sl Green}), and F658N ({\sl Blue}); (b) {\sl
Spitzer} 3.6 ${\rm \mu}$m ({\sl Red}), {\sl HST} H$\alpha$ ({\sl
Green}), and {\sl Chandra} source-removed intensity in the 0.5-1.5
keV band ({\sl Blue}). (c) {\sl Spitzer} 24~${\rm \mu}$m ({\sl
Red}), 3.6~${\rm \mu}$m ({\sl Green}), and
stellar-contribution-subtracted 8~${\rm \mu}$m (PAH, {\sl
Blue}).}\label{fig:tricolorimgs}
\end{center}
\end{figure}

We present the \chandra\ ACIS-S intensity images of \xs\ in
Fig.~\ref{fig:pointsrc} (a and b). In particular, the morphology of
the source-removed emission appears substantially rounder than that
of the stellar bulge (Fig.~\ref{fig:pointsrc}b), clearly indicating
the presence of non-stellar diffuse X-ray sources around the galaxy,
probably hot gas. The emission is further compared to the optical
and IR images in Fig.~\ref{fig:tricolorimgs}. The dusty disk is
prominent in the optical images as absorption feature, with a
diameter of $\sim1^\prime$ ($3\rm~kpc$), less extended than the
stellar disk which is also presented in the optical images
(Fig.~\ref{fig:tricolorimgs}a). The dusty disk is slightly tilted
from the stellar disk at the northwestern edge. In addition to the
prominent dust lane, there are also lots of dusty spurs evidenced as
extinction features in the optical images
(Fig.~\ref{fig:tricolorimgs}a). Diffuse soft X-ray emission is found
primarily in the bulge, while the H$\alpha$ emission is strongest
just off the dust lane (Fig.~\ref{fig:tricolorimgs}b). PAH and
24~$\rm \mu m$ emission is concentrated in the disk
(Fig.~\ref{fig:tricolorimgs}c), in similar fashion as the extinction
materials. But there is also a bright core on the 24~$\rm \mu m$
image (Fig.~\ref{fig:tricolorimgs}c), probably a nuclear SF region
or a weak AGN.

\begin{figure}[!htb]
\begin{center}
\centerline{
    \epsfig{figure=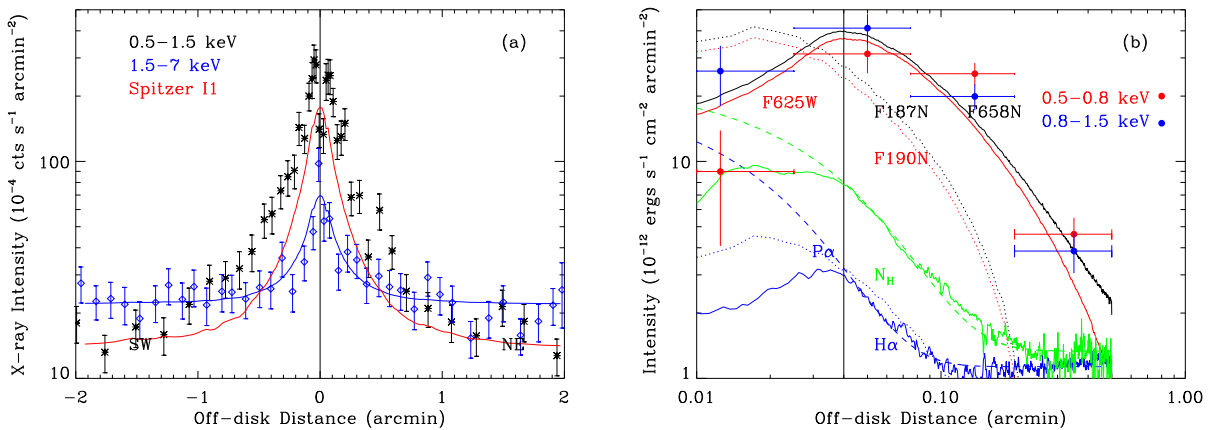,width=1.0\textwidth,angle=0, clip=}
} \caption{(a) Source-removed soft X-ray and near-IR intensity
profiles along the minor axis. The full width along the major axis
of the disk used for averaging the intensity is 3$^\prime$
($9.7\rm~kpc$). The {\sl Spitzer} IRAC 3.6~${\rm \mu}$m intensity
profile (blue solid line) is normalized to fit the 1.5-7~keV
profile. The red solid line shows the old stellar contribution in
0.5-1.5~keV band, calculated using the {\sl Spitzer} IRAC 3.6 ${\rm
\mu}$m intensity (see \S\ref{sec:hotgas} for details). (b) The {\sl
Chandra} source-removed X-ray and {\sl HST} optical intensity
profiles calculated for the central region (see the text for
details). The data on both sides of the major axis are averaged to
increase the signal-to-noise ratio. The full width along the major
axis of the disk used to average the intensities is 0\farcm5
($1.6\rm~kpc$) for all the data in (b). The $N_H$ profile calculated
from the F435W and F555W data is plotted with an arbitrary
normalization. This profile is only used to get the cold gas density
scale height (see \S\ref{sec:coolgas} for details). Both the
H$\alpha$ and $N_H$ profiles are fitted with an exponential model
(the dashed green and blue lines) in regions with
$|z|\gtrsim2\farcs4$ ($130\rm~pc$, marked by the black vertical
solid line, which marks the disk region where extinction is
substantial).}\label{fig:multiprof}
\end{center}
\end{figure}

Fig.~\ref{fig:multiprof} gives a more quantitative comparison of
these various intensity distributions along the minor axis of the
galaxy. It is clear that optical and soft X-ray intensities drop
significantly toward the galactic plane within an off-plane distance
of $|z|\sim2\farcs4$ ($130\rm~pc$, assume a distance of 11.21~Mpc as
listed in Table~\ref{table:basicpara} throughout the paper), due to
the extinction and absorption by dusty gas in the galactic disk. We
define distinct features outside this characteristic distance
($|z|\sim2\farcs4\sim130\rm~pc$) as "extraplanar". Assume the
diameter of the dusty disk to be $\sim1^\prime$ ($3\rm~kpc$) and
adopt the inclination angle in Table~\ref{table:basicpara}, the
projected vertical extension of the dusty disk is $\sim3\farcs6$
($200\rm~pc$), so the region with $|z|\gtrsim1\farcs8$ ($100\rm~pc$)
is out of the dusty disk, comparable to our definition of the
extraplanar region. In the following we detail the analysis and
results from the data sets.

\subsection{Optical and Infrared Properties}\label{sec:coolgas}

\begin{figure}[!htb]
\begin{center}
\centerline{
    \epsfig{figure=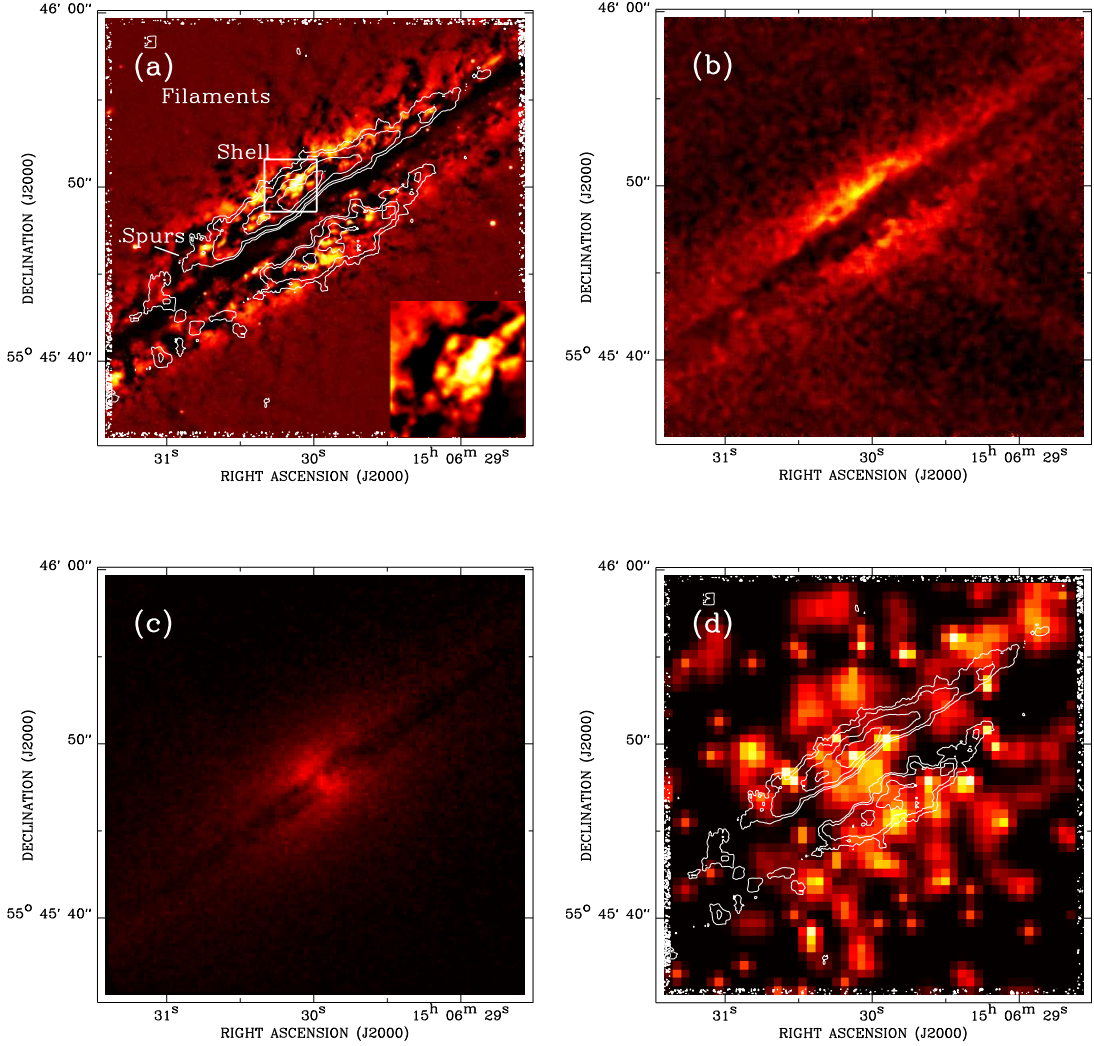,width=0.9\textwidth,angle=0, clip=}
} \caption{Multi-wavelength images of the central
$0\farcm4\times0\farcm4$ ($1.3\times1.3\rm~kpc$) of NGC~5866. (a)
H$\alpha$ contours overlaid on the unsharp-masked F435W image.
Various extinction features are marked. The white box encloses the
central shell-like feature, which is viewed in a close-up at the
lower right corner. (b) Smoothed H$\alpha$ image as in (a). (c)
Smoothed P$\alpha$ image. (d) H$\alpha$ contours as in (a) overlaid
on the smoothed diffuse 0.5-1.5 keV image.}\label{fig:multiimg}
\end{center}
\end{figure}

\begin{figure}[!htb]
\begin{center}
\centerline{
    \epsfig{figure=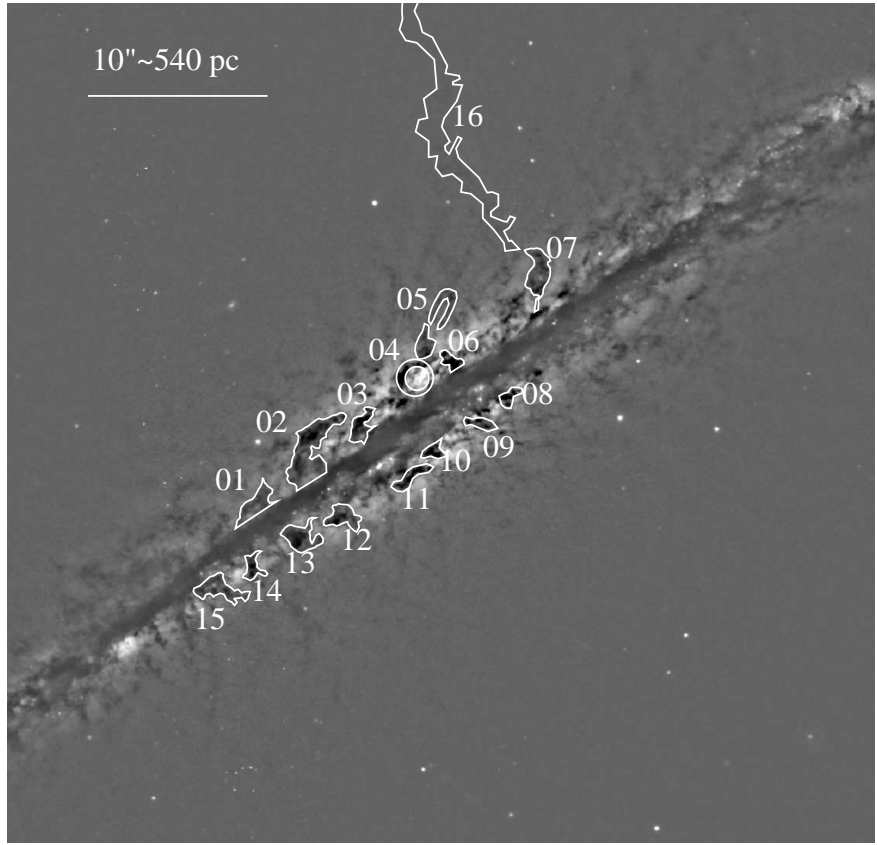,width=0.7\textwidth,angle=0, clip=}
} \caption{The unsharp-masked {\sl HST} F435W image. Individual dust
features are marked with numbers as listed in
Table~\ref{table:dustfeature}. The $10^{\prime\prime}$ bar shows the
linear and angular scale of the image.}\label{fig:dustfeature}
\end{center}
\end{figure}

We use the optical and infrared images to characterize the
interstellar extinction as well as the cold and warm gas components
in NGC~5866 (Figs.~\ref{fig:pointsrc}b, \ref{fig:tricolorimgs}).
Fig.~\ref{fig:multiimg} presents multi-wavelength images of the
central region of the galaxy. An unsharp-masked (Howk \& Savage
1997) B-band image is constructed from the {\sl HST} ACS F435W
image. It is first smoothed with a median filter of a
${51\times51~\rm pixel}~
({2\farcs5\times2\farcs5\sim135\times135\rm~pc})$ box and is then
subtracted from the original image. This unsharp-masked image could
emphasize structures on scales smaller than the smoothing box. As
shown in Figs.~\ref{fig:multiimg}a and ~\ref{fig:dustfeature}, it
highlights the filamentary structure of the extinction. In addition
to the prominent central dust lane along the galactic disk, various
prominent dusty spurs are located typically within $\sim0\farcm1$
($3\times10^2\rm~pc$) from the galactic plane. There is a shell-like
dusty feature near the nucleus, with a diameter of $\sim1\farcs8$
($\sim10^2\rm~pc$) (Fig.~\ref{fig:multiimg}a; i.e., Feature "04" in
Fig.~\ref{fig:dustfeature}). The unsharp-masked image also shows
more extended, but weaker extinction features in regions farther
away from the mid-plane (up to $\sim\rm kpc$
(${1\rm~kpc}\sim0\farcm3$) scale; e.g., Feature~"16" in
Fig.~\ref{fig:dustfeature}).

For a distinct small-scale feature, we may estimate its extinction
locally. The extinction at a particular wavelength $\lambda$ can be
expressed as
\begin{equation}\label{equ:apparentextinction}
a_\lambda=-2.5~log(S_{s,\lambda}/S_{b,\lambda})
\end{equation}
where $S_{s,\lambda}$ and $S_{b,\lambda}$ are the intensities on and
off the feature (see also Howk \& Savage 1997). This estimation
assumes that all the stellar light arises behind the feature, which
is reasonable for the stellar bulge with a centrally concentrated
light distribution. But in general, $a_\lambda$ should be considered
as a lower limit to the true extinction $A_\lambda$. We
adopt the relation between the total neutral hydrogen column density
and the color excess: $N_H=5.8\times
10^{21}~E(B-V)\rm~atoms~cm^{-2}$ (Bohlin, Savage \& Drake 1978),
assuming $R_V\equiv A_V/E(B-V)\approx 3.1$ and an extinction curve
to convert $A_\lambda$ to $A_V$ (Cox 1999). We further assume that
the depth of a dust feature is similar to its width and that the
hydrogen to total mass conversion factor is 1.37, appropriate for
gas of solar abundance. The resultant estimates of the hydrogen
column density, number density and total mass are presented in
Table~\ref{table:dustfeature}. A typical spur has mass
of $\sim10^4-10^5\rm~M_\odot$.  We estimate that
the mass of the extraplanar cold
gas is $\sim10^6-10^7\rm~M_\odot$ in total, much smaller than the total
cold gas mass estimated from CO
observations ($\sim5\times10^8\rm~M_\odot$,
Table~\ref{table:basicpara}). Thus most of the cold gas must have a
smoother distribution, concentrated in the disk.

To characterize the global extinction as a function of the vertical
distance $z$, we use the B-V color calculated with the {\sl HST}
F435W and F555W images. The optical color cannot be an exact measure
of the absorption column density when the intrinsic color is unknown
or in an optical thick case. Here we just use the color to show the
general distribution of the dusty gas and to give a crude estimate
of the vertical scale height of the cold gas. As described above,
$N_H\propto~E(B-V)$, so it can be written as $N_H\approx a\times
log\frac{S_V}{S_B}-b$, where $S_V$ and $S_B$ are the V- and B-band
surface brightness, $a\approx1.5\times10^{22}\rm~cm^{-2}$, $b$ is a
parameter determined by the intrinsic color of the stellar
population. Assuming a constant intrinsic color, the cold gas
distribution is determined by the term $log\frac{S_V}{S_B}$, which
is plotted in Fig.~\ref{fig:multiprof}b and is fitted with an
exponential model for the extraplanar region with
$|z|\gtrsim2\farcs4$ ($130\rm~pc$). The scale height of the cold gas
so obtained is $\sim10^2\rm~pc$. The flatten of the $N_H$ profile in
the central region with $|z|\lesssim2\farcs4$ ($130\rm~pc$) is due
to the tilt of the disk, that is why we have defined regions with
$|z|\gtrsim2\farcs4$ ($130\rm~pc$) as "extraplanar". The innermost
trough of the $N_H$ profile is caused by the extinction saturation
in B-band.

Warm ionized gas, as traced by the H$\alpha$ or P$\alpha$ emission,
shows a relatively smooth morphology and extends to $|z|\sim
300\rm~pc$ ($\sim0\farcm1$) from the mid-plane (e.g.,
Fig.~\ref{fig:multiprof}b, Fig.~\ref{fig:multiimg}b,c). Although its
overall distribution is similar to that of the dusty cold gas, there
is little obvious feature-to-feature correlation. Indeed, some of
the dusty features appear in regions with weak H$\alpha$ emission
(e.g., Feature "02" in Fig.~\ref{fig:dustfeature}). We fit the
vertical distribution of the H$\alpha$ emission, instead of
P$\alpha$ (because of the small field coverage of the NICMOS), in
the region with $|z|\gtrsim2\farcs4$ ($130\rm~pc$)
(Fig.~\ref{fig:multiprof}b). The fitted exponential scale height is
$\sim58\rm~pc$. Since the H$\alpha$ intensity is proportional to
$n_{warm}^2$, the scale height of the warm gas density should be
$\sim10^2\rm~pc$, comparable to the scale height of the optical
absorption features (or cold gas).

As the H$\alpha$ emission suffers stronger extinction than the
P$\alpha$ emission, we use the P$\alpha$ luminosity to estimate the
recombination rate of the warm ionized gas. The total observed
P$\alpha$ luminosity in the region represented by the second X-ray
data point at $|z|\sim3''$ ($160\rm~pc$) (Fig.~\ref{fig:multiprof}b)
is $(0.7-1.5) \times 10^{39}\rm~ergs~s^{-1}$, accounting for the
uncertainty in the normalization factor of the continuum subtraction
(\S\ref{sec:data}) and corrected for the extinction. This luminosity
corresponds to a recombination rate of
$(4.3-9.6)\times10^{51}\rm~s^{-1}$.

\subsection{X-ray Properties}\label{sec:hotgas}
%here not only hot gas is described.

We detect 15 point sources within the bulge region
(Fig.~\ref{fig:pointsrc}). In particular, Source 30 coincides with
the optical center of the galaxy (Fig.~\ref{fig:pointsrc}c). This source,
detected only in the 1.5-7~keV band, has a count rate of $0.61\rm~cts~ks^{-1}$
(Table~\ref{table:acis_source_list}). Assuming a typical
power law spectrum with a photon index of 1.7, we estimate
the corresponding 1.5-7~keV
luminosity as $\sim 1.2\times10^{38}\rm~ergs~s^{-1}$. The apparent
strong soft X-ray absorption is consistent with the nucleus being behind the
dusty disk. In any case, the low luminosity of the nucleus indicates
that it is in a relatively quiescent state and probably has little
impact mechanically on the global ISM of the galaxy. The spectrum of
other sources outside the disk (Fig.~\ref{fig:pointsrc}b),
the bulk of which should be LMXBs in the
galactic bulge, can be represented by a power law of a photon index
$\Gamma\approx1.5$ and a very low absorption, consistent with the
Galactic foreground absorption value. Their integrated intrinsic
luminosity is $\sim3.8\times10^{38}\rm~ergs~s^{-1}$ in the 0.5-2~keV
band.

By removing the detected sources, we study the apparent diffuse
emission. To isolate the truly diffuse hot gas in the galaxy,
however, we need to quantify remaining stellar contributions: the
emission from cataclysmic variables (CVs) and coronal active
binaries (ABs) as well as the emission from unresolved LMXBs and the
residual photons out of the discrete source removal circles
(\S\ref{sec:data}). We estimate the CV and AB contribution,
$2.1\times10^{38}\rm~ergs~s^{-1}$, from the calibrated ratio between
the 0.5-2~keV luminosity and the bulge stellar mass
($\sim7.0\times10^{27}\rm~ergs~s^{-1}~M_\odot^{-1}$; Revnivtsev et
al. 2008; refer to \S\ref{subsec:SF} for the estimation of the
stellar mass). The contributions in other bands are calculated with
a spectral model consisting of a 0.5~keV {\sl MEKAL} optically-thin
thermal plasma and a $\Gamma\approx1.9$ power law (Revnivtsev et al.
2008). The residual emission from the source removal is $\sim4\%$ of
the detected source luminosity. The emission from unresolved LMXBs
below our detection limit ($\sim4\times10^{37}\rm~ergs~s^{-1}$ in
the 0.5-8~keV band) is estimated from the luminosity function of
Gilfanov (2004), assuming the same spectrum as that for the detected
sources. We then find that the total residual LMXB contribution is
$\sim3.3\times10^{38}\rm~ergs~s^{-1}$ in 0.5-2~keV band. This
contribution is confirmed independently from the spectral analysis
of the diffuse emission, in which the normalization of the residual
LMXB contribution is fitted (the luminosity of this component is
fitted to be $\sim2.9\times10^{38}\rm~ergs~s^{-1}$ in 0.5-2~keV
band, see below). These stellar contributions are added together and
are assumed to have the same spatial distribution as the IRAC
$3.6\rm~\mu m$ intensity of the galaxy. They are then subtracted
from the unresolved X-ray emission in spatial
(Fig.~\ref{fig:multiprof}a) and spectral (see below) analysis.

Fig.~\ref{fig:multiprof}a compares the intensity distributions of
the stellar contributions with the actually observed diffuse X-ray
emission in the 0.5-1.5~keV and 1.5-7~keV bands. The two data points
at the galactic plane in Fig.~\ref{fig:multiprof}a shows large
deviations from the global distribution of the 0.5-1.5~keV intensity
profile. These deviations are clearly due to the absorption of soft
X-ray by the cold gas disk. The stellar contribution as traced by
near-IR emission matches well the observed 1.5-7~keV emission,
indicating a stellar origin of the hard X-ray emission. Using the
stellar contribution in 1.5-7~keV band as fitted and the spectral
model of stellar sources adopted in the X-ray spectral analysis, we
calculate the stellar contribution in 0.5-1.5~keV band (the red
solid line in Fig.~\ref{fig:multiprof}a). The excess of the observed
0.5-1.5~keV emission above the stellar contribution clearly shows
the presence of the truly diffuse hot gas widespread in the galactic
bulge, with a luminosity of $\sim6.5\times10^{38}\rm~ergs~s^{-1}$ in
a region with
$2\farcs4~({130\rm~pc})\lesssim|z|\lesssim0\farcm55~({1.8\rm~kpc})$
(roughly the bulge region shown in Fig.~\ref{fig:pointsrc}).

%[Again, if X-ray emission from hot gas depends on the density square and is not expected to follow the stellar light distribution even if stars are responsible for the heating! The two distributions are not really that similar especially at large off-plane distancesi! Although has a rounder morphology (Fig.~\ref{fig:pointsrc}b), in general the diffuse soft X-ray emission shows similar distribution as the stellar light in the vertical direction (Fig.~\ref{fig:multiprof}a), indicates hot gas may be produced by processes related to old stars.

\begin{figure}[!htb]
\begin{center}
\centerline{
    \epsfig{figure=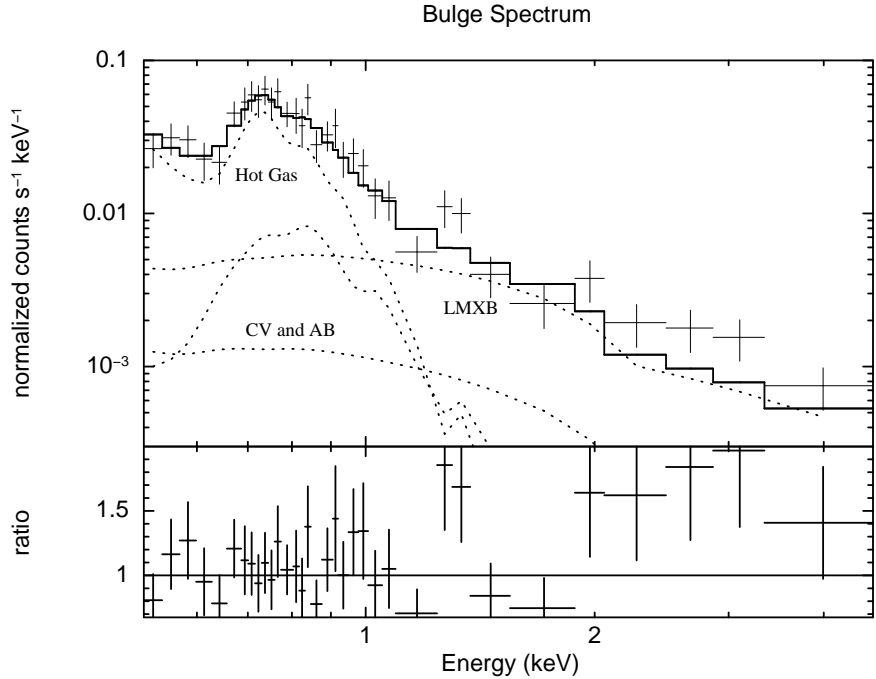,width=0.7\textwidth,angle=0, clip=}
} \caption{Spectrum of the diffuse X-ray emission from the bulge
region. This spectrum is binned to achieve a signal-to-noise ratio
better than 3 with respect to the best-fitted local background. It
is also sky background and instrumental background subtracted. The
best-fit model and its components (marked individually) are
indicated by solid and dotted curves. Part of the residual photons
from the model above 2~keV is due to some line features of the
instrument particle signals (which is filtered during the fitting),
although they are not clear on the figure, which is regrouped after
subtracting the instrumental background. See text for
details.}\label{fig:bulgespec}
\end{center}
\end{figure}

Fig.~\ref{fig:bulgespec} presents the spectrum of the diffuse X-ray
emission from the bulge, together with our best-fit model. In
addition to the inclusion of the stellar contributions as described
above, we model the hot gas emission as an optically-thin thermal
plasma ({\sl VMEKAL}). The fitted temperature of the plasma is
$0.17_{-0.03}^{+0.02}\rm~keV$. The Fe and O abundances are
$3.5_{-2.1}^{+1.4}$ and $\lesssim0.16$ solar. The inferred intrinsic
luminosity of the hot gas $\sim7.0\times10^{38}\rm~ergs~s^{-1}$ in
the 0.5-2~keV band is consistent with the above estimation based on
the spatial distribution. The corresponding emission measure of the
hot gas is $\sim0.07\rm~cm^{-6}~kpc^3$. Note that the exact value of
the metal abundance is somewhat uncertain and is model-dependent due
to the limited counting statistics and spectral resolution of the
spectrum, but the assumption of isothermality typically leads to an
under-estimate of the metal abundance. Therefore, the X-ray spectrum
strongly indicates supersolar iron abundance, consistent with
significant enrichment by Type~Ia SNe.

\section{Discussion}\label{sec:discussion}

Based on the above multi-wavelength characterizations of the stellar
and interstellar components of \xs, we now discuss their interplays, focusing
on the physical state, dynamics, and evolution of the extraplanar gas.
But first let us characterize the SF history of the galaxy,
important for estimating the stellar mass and energy feedback.

\subsection{Star formation history}\label{subsec:SF}

For an S0 galaxy, we may assume that the bulk of stars is formed
long ago during or before the last major merger that presumably
forms the galactic bulge. One way to estimate the age of this merger
is to use the globular cluster (GC) specific frequency ($S_N$), the
number of GCs per unit luminosity in a given galaxy. GCs are
believed to form in starbursts triggered by major mergers of
gas-rich galaxies. Assuming that the evolution after mergers
preserves the GC number, while the overall stellar luminosity of the
galaxy may decrease with time. Therefore, $S_N$ can be used as an
age estimate for an S0 galaxy after its last merger and
accompanied starburst (Barr et al. 2007). With $S_N\sim1.4$
(Cantiello, Blakeslee \& Raimondo 2007), we estimate the age of the
NGC~5866 bulge to be $\sim8\rm~Gyr$, which is typical for S0
galaxies (Barr et al. 2007).

We now estimate the stellar mass of NGC~5866. Using the \emph{2MASS}
J- and K-band images, we get the J- and K-band intensity
distributions. We further assume an optically thin case for near-IR
emission, and an intrinsic color J-K=0.90, suitable for old stars in
early type galaxies (Jarrett et al. 2003). We then map out the
extinction value in different parts of the galaxy. Using the
extinction-corrected K-band luminosity and adopting a
color-dependent stellar mass-to-light ratio as detailed in Bell \&
de~Jong~(2001), we estimate the masses of the galactic bulge and
disk regions (Fig.~\ref{fig:pointsrc}b) as
$\sim1.8\times10^{10}\rm~M_\odot$ and
$\sim1.1\times10^{10}\rm~M_\odot$.

There are lines of evidence for ongoing SF in the disk of \xs. The
most direct evidence comes from the observed 24~$\rm\mu m$ emission,
which is usually thought to be mainly produced by hot dust heated by
ultraviolet radiation, and is observed to concentrate in the cold
gas disk in \xs ~(Fig.~\ref{fig:tricolorimgs}c). Such hot dust is
found in both SF regions and circum-stellar material around evolved
stars. The latter contribution, tightly correlated with the K-band
luminosity in elliptical and S0 galaxies (e.g., Temi, Brighenti \&
Mathews 2007, 2009), is estimated to be
$\sim7\times10^{40}\rm~ergs~s^{-1}$ for \xs, significantly smaller
than the total observed 24~$\rm\mu m$ luminosity of
$\sim3\times10^{41}\rm~ergs~s^{-1}$ (excluding the nucleus which may
be related to the AGN). This excess of
$2.3\times10^{41}\rm~ergs~s^{-1}$ predicts an ongoing SF rate of
$\sim0.05\rm~M_\odot~yr^{-1}$ in the galactic disk. Here we have
assumed that the relationship, SFR $({\rm M_\odot~yr^{-1}})= 1.27
\times 10^{-38} [L_{24 {\rm~\mu m}} ({\rm~ergs~s^{-1}})]^{0.8850}$,
obtained for active star forming galaxies (Calzetti et al. 2007),
applies to NGC~5866. The low level of SF inferred here is consistent
with the constraint ($< 0.1 {\rm~M_\odot~yr^{-1}}$) obtained from
the model fit to the observed SED of NGC~5866 over the
$\sim2-10^3\rm~\mu m$ range (Draine et al. 2007). This means the SFR
estimated using the mid-IR emission is reliable within the errors
for NGC~5866, and further indicates that the MIPS 24~$\rm\mu m$
luminosity is mainly produced by the weak SF process in the disk.

The ongoing SF also explains the
ionizing photon flux needed for producing the P$\alpha$ luminosity.
The relationship
$L_{P\alpha}=0.0049~L_{24~\rm\mu m}$ (Rieke et al. 2009) gives a
P$\alpha$ luminosity of $1.1\times10^{39}\rm~ergs~s^{-1}$, fully
consistent with the observed value, $(0.7-1.5)\times10^{39}\rm~ergs~s^{-1}$,
which corresponds to a recombination rate of
$(4.3-9.6)\times10^{51}\rm~s^{-1}$ (\S~\ref{sec:coolgas}).
This may be a lower limit, since we may have still missed part of the
P$\alpha$ emission from the disk, due to its strong extinction.
Alternatively, could the ionizing radiation be supplied by
evolved stars? We use the ionizing radiation with a rate of
$7.3\times10^{40}\rm~s^{-1}~M_\odot^{-1}$ for post-AGB stars
(Binette et al. 1994) and $1.4\times10^{40}\rm~s^{-1}~M_\odot^{-1}$
for hot horizontal branch stars (Han et al. 2007). Adopting the
total stellar mass of the galaxy, $\sim3\times10^{10}\rm~M_\odot$ as
inferred in \S\ref{sec:coolgas}, we estimate that the total expected
hydrogen recombination rate for the galaxy is
$\sim3g_\star\times10^{51}\rm~s^{-1}$, where $g_\star$ is the
fraction of the evolved ionizing photons intercepted by the gas.
While most of the stars are in the
galactic bulge, $g_\star$ should be considerably smaller than one.
Therefore, the evolved stars alone seem to be insufficient to provide the
ionizing photons. Furthermore, observed various emission lines such
as [\ion{C}{2}]$\lambda158\rm\mu m$ and [\ion{O}{1}]$\lambda63\rm\mu
m$ appear to be considerably stronger than what are expected from
ionizing photons purely from the evolved stars (Malhotra et al. 2000, 2001).

Therefore, we conclude that the presence of low-level SF provides a
unified explanation for the radiative heating and ionization of the
dust and warm-ionized gas in \xs.

\subsection{Energetics of the extraplanar hot gas}\label{sec:ISM}

Hot gas apparently dominates the extraplanar space. Adopting the
spectral fit parameters in \S\ref{sec:hotgas} and approximating the
X-ray-emitting region as a cylindric volume with the radius
$\sim1\farcm3$ ($4.2\rm~kpc$) and the hight $\sim0\farcm55$
($1.8\rm~kpc$) (Fig.~\ref{fig:pointsrc}b), we estimate the mean
density, thermal pressure, and radiative cooling time scale of the
hot gas as $\sim0.065\rm~cm^{-3}$, $\sim10^{-11}\rm~dyne~cm^{-2}$,
and $\sim1.5\times10^8\rm~yr$. The total hot gas mass is
$\sim3\times10^7\rm~M_\odot$. The estimation of these parameters are
all based on the most reliable spectral fit parameters, i.e., the
temperature, the luminosity and the emission measure of the thermal
component (\S\ref{sec:hotgas}).

The most natural source of the gas that is subsequently heated is
the mass loss of evolved stars. Assuming the mass loss by each
planetary nebula (PN) is $\sim0.3\rm~M_\odot$ (Faber \& Gallagher
1976), and using the PN birth rate in our Galaxy
$\sim11\times10^{-12}\rm~yr^{-1}~L_\odot^{-1}$ (Peimbert 1993), the
mass loss rate ($\rm0.8~M_\odot<M<8~M_\odot$) is
$0.033\times10^{-10}\rm~M_\odot~yr^{-1}~L_\odot^{-1}$. Adopting the
stellar mass estimated in \S\ref{subsec:SF} and the age of
$\sim8\rm~Gyr$ (\S\ref{subsec:SF}), the total mass return in
NGC~5866 is then $\sim8\times10^8~M_\odot$, which is substantially
greater than what is observed in the hot gas. Furthermore, the
mass-averaged Fe abundance as estimated from the mass loss of the
evolved stars and metal enrichment of the Type Ia SN ejecta is about
6 times the solar value (Grevesse \& Suval 1998), much larger than
our spectral fit result (\S\ref{sec:hotgas}). One possibility of
this relatively low observed Fe abundance is that the bulk of the
Fe-rich SN ejecta is being blown out of the bulge before being fully
mixed with the materials from the stellar mass loss (Tang et al.
2009).

To see if the heating provided by Type~Ia SNe is sufficient to
explain the X-ray cooling luminosity, we compare the X-ray
luminosity with the SN heating rate. The supersolar iron abundance,
as inferred from the spectral analysis of the hot gas, is consistent
with heating primarily by Type~Ia SNe. Assuming that each SN
releases $10^{51}\rm~ergs$ and using the empirical Type~Ia SN rate
of Mannucci et al. (2005), which can be approximately expressed as
$4.4\times10^{-4}~(M/10^{10}\rm~M_{\odot}) {\rm~yr^{-1}}$, and the
stellar mass $\sim2.9\times10^{10}\rm~M_\odot$ (\S\ref{subsec:SF}),
we estimate the total Type~Ia SN rate of
$1.3\times10^{-3}\rm~SN~yr^{-1}$ and the mechanical energy injection
rate as $\dot{E}_{SNIa}\sim 4\times10^{40}\rm~ergs~s^{-1}$. This
energy injection rate is nearly two orders of magnitude greater than
the X-ray luminosity! While a fraction of the energy injected into
the cool gas disk may be radiated in other energy bands (e.g.,
extreme- and far-UV), avoiding a detection so far, the general lack
of cool gas in the galactic bulge makes it hard to hide and convert
the mechanical energy. Therefore, there is a missing stellar
feedback problem in NGC~5866, similar to situations in other
low-to-intermediate mass early-type galaxies (e.g. Li, Wang \&
Hameed 2007; Li, \& Wang 2008; David et al. 2006). This missing
stellar feedback indicates the presence of an outflow from the
galactic bulge.

The dynamics of this outflow depends critically on the mass-loading
from the cool gas disk, which is indicated by the observed low
temperature of the hot gas in \xs. The specific enthalpy of the
stellar feedback, $\sim4\rm~keV$ per particle, would give an average
gas temperature of $\sim3\rm~keV$ at the bulge center, which is
substantially greater than $\sim0.17\rm~keV$ inferred from the X-ray
spectrum of the bulge (\S\ref{sec:hotgas}). In the inner bulge, the
kinetic energy is often negligible compared to the thermal energy of
the hot gas (Tang et al. 2009), so it cannot account for this large
temperature difference. Part of the difference can be attributed to
the SN energy stored in other forms (possibly magnetic field and
cosmic rays as well as turbulent motion), the equal-partition among
which may account for a factor up to 3-4. The emission
measure-weighted temperature, as in the X-ray spectral analysis,
also tends to under-estimate the real gas temperature by a factor
$\sim 2$ (Tang et al. 2009). At least an additional factor of 2-3 is
needed to explain our measured low temperature. This factor can be
naturally explained by the mass-loading from the cool gas disk, as
evidenced by the presence of extraplanar warm-ionized gas and dusty
spurs, which represent the interface between the hot gas in the
bulge and the cool gas in the disk.

\subsection{Energetics of the extraplanar cool gas}\label{sec:coolgasenergetics}

Assuming a temperature of $\sim10^4\rm~K$ and an approximate
pressure balance with the hot gas, we estimate the parameters of the
warm ionized gas based on the observed P$\alpha$ emission measure.
The volume filling factor, number density and total mass of the warm
gas estimated this way are $\sim10^{-4}$, $10\rm~cm^{-3}$ and
$10^4\rm~M_\odot$. The total mass is roughly consistent with the
value of $4\times10^4\rm~M_\odot$ as listed in
Table~\ref{table:basicpara}. However, the number density is too high
compared to that in other nearby galaxies (e.g. Collins et al.
2000), indicating that the warm and hot gas may be not in a pressure
balance state, or the warm ionized gas may present as skins of some
unresolved cool gas filaments.
%Of course, there is likely a more diffuse component of the warm gas, which may also contribute significantly to the observed H$\alpha$ and P$\alpha$ emission.

Similar dusty spurs and their correlation with the extraplanar
ionized gas have been observed previously in nearby edge-on {\sl
active star-forming} galaxies (e.g., Howk \& Savage 1999; Li et al.
2008; Rossa et al. 2008). In particular, Howk \& Savage (1997) have
presented several mechanisms for the formation of extraplanar dusty
features observed in NGC~891. These mechanisms include
hydrodynamical processes like galactic fountain, radiation pressure
driven outflow, and magnetic instability presumably induced by the
differential rotation of galactic gaseous disks. The galactic
fountain, the radiation pressure, and some magnetic instabilities
like the Parker instability are closely related to recent star
forming activities. This is consistent with an apparent correlation
between the amount of extraplanar gas and the SF rate of galaxies
(Lehnert \& Heckman 1996).

%The detection of the extraplanar dusty spurs in NGC~5866 provides new insight into their formation.
The heights and masses of the extraplanar cold gas features in
NGC~5866 are significantly smaller than that in late-type galaxies
(Howk \& Savage 1999). Thus the processes responsible for producing
the features are probably less energetic and would be overwhelmed in
an active star forming galaxy. The very detection of the extraplanar
extinction features in NGC~5866 is largely due to its nearly perfect
edge-on perspective and its prominent stellar bulge background
light.
%While magnetic instabilities may play a role in developing the disk/halo interaction, other known energy sources such as Type Ia SNe must also be considered.
To check out the mechanical energy requirement for the
extraplanar extinction features, we estimate their gravitational
potential energies. Following Howk \& Savage (1997), we estimate the
energy $\Omega$ of a feature at height $z$ from the mid-plane as:
\begin{equation}\label{equ:potential}
\Omega=10^{52}~(\frac{M_c}{10^5~{\rm M_\odot}})(\frac{z_0}{300~{\rm
pc}})^2(\frac{\rho_0}{1~{\rm
M_\odot~pc^{-3}}}){\rm ln[cosh}(z/z_0)]\rm~ergs
\end{equation}
where $M_c$ is the mass of the cold gas feature, $z_0$ is the scale
height of the mass density and is taken to be $\sim300\rm~pc$
($\sim0\farcm1$), and $\rho_0\sim8.5\rm~M_\odot~pc^{-3}$ is the mass
density at the mid-plane derived from the mass of the stellar disk
(\S\ref{subsec:SF}) and its size. The estimated gravitational
potential energies for individual spurs are included in
Table~\ref{table:dustfeature}.

The mechanical energy needed to overcome the gravitational potential
energy for an individual spur is mostly comparable to that expected
for a single SN. Some of the spurs may represent super-impositions
of multiple features. In such cases, the required energy could be
considerably less. So in principle such spurs can be produced by
individual SNe (e.g., Raley, Shelton \& Plewa 2007), although some
may be due to concerted energy inputs from massive stars, which may
be formed in small groups or associations in \xs. Using
the empirical SFR-dependent relationship of Cappellaro,
Evans \& Turatto~(1999), we estimate the core collapsed
SN rate as $5\times10^{-4}\rm~SN~yr^{-1}$, which is
smaller than the Type~Ia SN rate of $1.3\times10^{-3}\rm~SN~yr^{-1}$
as estimated in \S\ref{sec:ISM}. Thus in NGC~5866,
Type~Ia SNe may be the primary energy source of the
extraplanar dusty spurs, although core collapsed SNe can also be important.
%Thus the galactic fountain and radiation pressure should not be important.
In contrast, the required energy for a
high-$z$ cold gas feature in late-type galaxies
typically equals tens or hundreds of the
SN energy (Howk \& Savage 1997, 1999; Thompson, Howk \& Savage
2004). This large energy requirement means most of the high-$z$ cold
gas features in these galaxies should be produced by the concerted energy
injection from massive star forming regions. Similarly, large scale
cold gas filaments (e.g., Feature~"16" in
Fig.~\ref{fig:dustfeature}) detected in NGC~5866 also require
substantially more energy to form. They are probably produced by
mechanical energy inputs provided collectively by core collapsed SNe
in the disk and/or Type~Ia SNe in the bulge of the galaxy.

\subsection{Gas circulation between the galactic disk and halo }

We further estimate the disk mass loss rate via outflows as
indicated by the extraplanar dusty spurs in \xs. From the simulation
of extraplanar SNe driven disk outflow (Raley, Shelton \& Plewa
2007), the typical timescale to form an extraplanar cold gas spur
through SNe driven hydrodynamical processes is $\sim 10^7\rm~yr$.
From the mass of the observed extraplanar cool gas ($\sim5 \times
10^6\rm~M_\odot$; \S\ref{sec:coolgas}), we infer the disk mass loss
rate as $\sim0.5\rm~M_\odot~yr^{-1}$. A mass-loading to the hot
phase at a comparable rate can naturally explain the low temperature
of the hot gas (\S\ref{sec:ISM}). But, assume a constant mass-loss
rate of $\sim0.5\rm~M_\odot~yr^{-1}$, the disk would have lost $\sim
4\times10^9\rm~M_\odot$ over 8~Gyr, which is an order of magnitude
more than what remains in the disk. The current mass loss from the
disk is driven by individual SNe (\S\ref{sec:coolgasenergetics}), a
scenario different from that assumed for more energetic features
detected in more active star forming galaxies. However, during the
evolution of the galaxy, different processes (e.g. Type~Ia SNe and
SF) may dominate in different stages, so the mass loss rate may not
be constant. In addition, accretion of cold gas from the
intergalactic space could to some degree replenish the mass loss
from the cold gas disk, which will also affect the observed amount
of cold gas in the disk.

Some of the outflowing hot gas will eventually cool and return to
the disk. Indeed, the mass-loaded hot gas with a temperature of
$\sim 0.17$~keV is not expected to escape the gravitational
potential of the galaxy. The corresponding sound speed of the hot
gas is only about $c_s\simeq150{\rm~km~s^{-1}}$, which is
considerably smaller than the escape velocity, which should roughly
be about three times of the circular velocity
($\sim265\rm~km~s^{-1}$; Neistein et al. 1999)
%; Table~\ref{table:basicpara})
for a typical dark halo potential. In this case, the hot gas will be
confined by the gravitational potential of the galactic halo. The
outflow is likely to be subsonic, explaining the moderate X-ray
luminosity of the hot gas emission from \xs\ (Tang et al. 2009). The
upper limit to the outflow speed, $v$, may be estimated from the
accumulation time scale ($\gtrsim5\times10^7$~yrs) of the hot gas in
the bulge via the mass loss of evolved stars and from the cool gas
mass loading from the disk. The lower limit can be set to have the
cooling time scale ($\sim1.5\times10^8$~yrs) of the hot gas longer
than its escape time scale, assuming a characteristic bulge X-ray
emission radius to be $\sim 3$ kpc. We thus have $0.4 c_s \gtrsim v
\gtrsim 0.1 c_s$. The gas will cool in the large-scale galactic halo
and will eventually return to the disk/bulge of the galaxy, which
may be partly responsible for the present dusty galactic disk.
Therefore, we expect a gas circulation, or a halo fountain (i.e., an
outflow of heated gas to the halo and an inflow of cool gas to the
disk/bulge), some of the extraplanar dusty structures may be such
cool gas inflow (Fig.~\ref{fig:dustfeature}).

\subsection{Formation and evolution of field S0 galaxies}\label{sec:gasorigin}

Now let us see how \xs\ may fit into the existing picture of galaxy
formation and evolution. S0 galaxies in general are believed to be
evolved from spirals when they have exhausted gas for massive SF.
Much of the existing work has been focused on S0 galaxies in cluster
environments, where the exhaustion of the gas is typically
attributed to the ram-pressure stripping by the ambient medium (e.g.
van~Dokkum et al. 1998; Bekki, Couch \& Shioya 2002; Goto et al.
2003; Moran et al. 2006). Outside clusters, the ram-pressure is
unlikely to be strong enough to remove cool gas from galactic disks,
although it may still be important in stripping gas from extended
galactic halos and/or regulating the accretion from the
intergalactic medium (IGM; e.g., Kawata \& Mulchaey 2008). The
evolution of such a field galaxy at present is typically
dominated by internal processes, with galactic mass as the key
parameter, according to recent simulations of galaxy formation
(Birnboim \& Dekel 2003; Birnboim et al. 2007; Kere\v{s} et al.
2005, 2009). A galaxy with $M_{halo}\gtrsim10^{11}\rm~M_\odot$ is
expected to obtain its gas primarily in a so-called hot mode, in
which accreted gas gets shock-heated near the virial radius to $T
\gtrsim 10^6$~K. This virial shock presumably does not develop in
lower mass galaxies, and the accretion is in a cold mode, in which
the gas flows along narrow filaments that extend well inside the
virial radius. This simple picture may provide a natural
interpretation for the observed galaxy bi-modality of stellar color
and morphology, although complications arise when galaxy-galaxy
mergers and galactic feedback are considered. In this evolution
picture, S0 galaxies may be a direct result of their mass growth,
passing through the mass threshold, through mergers and/or
accretion.

However, a field S0 galaxy formed this way (shutting off the cool
gas supply) is expected to contain certain amount of cool gas in the
galactic disk. In addition to the accumulation of the stellar mass
loss, as discussed earlier, we also expect residual gas in the
galactic disk from past SF. The consumption via SF becomes
inefficient as the amount of cool gas slowly decreases (e.g.,
considering the Schmidt-Kennicutt Law and the Toomre column density
criterion of SF; Kennicutt~1998; Toomre~1964). So substantial
amounts of the disk gas should remain when the SF has more or less
ceased, which may be the case for NGC~5866 and other field S0
galaxies (e.g., Sage \& Welch 2006). Of course, as demonstrated in
the present work, the feedback from old stars in the galactic disk
and bulge, though relatively gentle, is long lasting, which can help
to evaporate the cool gas disk. The effectiveness of this process on
the long run depends on whether or not the evaporated gas can cool
and then fall back into the disk, which is in turn sensitive to the
intergalactic ram-pressure that the gaseous halo is subject to.
NGC~5866 is relatively rich in cool gas as an S0 galaxy (Welch \&
Sage 2003; Sage \& Welch 2006), which may be due to a lower than
average ram-pressure --- a result of the galaxy's relative isolation
in a low-density environment. How the environment and S0 galaxy
evolution affect the gas content and other galactic properties will
be discussed in a following paper.

\section{Summary}\label{sec:summary}

We have presented a multi-wavelength study of the edge-on S0 galaxy
NGC~5866 in a very low density environment.
%This galaxy has an unusual large amount of cold gas for a S0 galaxy.
We characterize various stellar and gaseous components and discuss
the galactic disk/halo interaction and its role in the evolution of
S0 galaxies. The main results are as follows:

\begin{itemize}
\item Based on a \chandra\ ACIS observation, we detect 15 point-like sources
within the galactic bulge of the galaxy. One of them, with a
1.5-7~keV luminosity of $\sim1.2\times10^{38}\rm~ergs~s^{-1}$,
probably represents a very weak AGN of the
galaxy.

\item After removing the detected sources and subtracting remaining
stellar contributions, we reveal the X-ray emission from the
diffuse hot gas in and around the galactic bulge. The emission
extends as far as 3.5 kpc away from the galactic plane and has a total
luminosity of $\sim7\times10^{38}\rm~ergs~s^{-1}$ in 0.5-2 keV.
With a characteristic
temperature of $\sim0.2\rm~keV$ and a total mass of
$\sim3\times10^7\rm~M_\odot$, the hot gas most likely represents a combination
of the expected stellar mass loss and the mass-loading
from the cool gas disk of the galaxy. The hot gas
is over-abundant in iron and under-abundant in oxygen, indicating a
strong chemical enrichment by Type~Ia SNe.

\item We detect warm ionized gas based on \hst\ observations
of H$\alpha$/P$\alpha$ emission. This gas has an exponential scale
height of $\sim10^2\rm~pc$ and extends as far as
$|z|\sim3\times10^2\rm~pc$ from the mid-plane, indicating the
presence of an extraplanar component. Massive stars in the galactic
disk are primarily responsible for the ionization of this warm
ionized gas.

\item We characterize numerous cold gas features off the
galactic disk, based on their extinction against the bulge stellar
light. Dusty spurs typically have extensions similar to the
warm ionized gas and have individual masses of
$\sim10^{4-5}\rm~M_\odot$. Such spurs may be generated mostly by individual
SNe in the galactic disk. But there are also more energetic
kpc-scale filaments, which may be produced by
SNe collectively in the disk and/or bulge of the galaxy.

\item We show strong evidence for ongoing massive SF in the disk of
the galaxy at a rate of $\sim0.05\rm~M_\odot~yr^{-1}$. The presence of
the massive stars is required for the
heating of the dust and warm ionized gas. The stars, which tend to
be formed in groups, may be responsible for some of the energetic
extraplanar cool gas features.

\item The relative richness of cool gas in the galactic disk is probably
related to the isolation of the galaxy in a low density environment.
This gas, while being slowly consumed by the SF, may be replenished
by the mass loss of evolved star, mostly in the galactic bulge.
The gas appears to be undergoing a
circulation between the cool and hot phases via the galactic disk/halo
interaction. This circulation becomes possible for the galaxy
because of both the low temperature of the mass-loaded hot gas and the
inefficiency of ram-pressure stripping in the environment.

\item A field S0 galaxy may form mainly from the development of a hot
gaseous halo, which stops or slows down the supply of cool gas from the IGM.
This development may be a natural consequence of the mass growth
of a galaxy, according to recent structure formation simulations.
But the inclusion of the galactic feedback as well as the environmental
effects, as indicated in the present work, may also be important.

\end{itemize}

\acknowledgements

This work is supported by NASA through the CXC/SAO grants AR7-8016A
and G08-9088B, by NSFC through the grants 10725312 and 10673003 and
the China 973 Program grant 2009CB824800, and by China Scholarship
Council.

\begin{deluxetable}{lrrrrrrrr}
  \tabletypesize{\footnotesize}
  \tablecaption{Basic Information of NGC~5866}
  \tablewidth{0pt}
  \tablehead{
  \colhead{Parameter} &
  \colhead{NGC~5866} &
    }
  \startdata
     Morphology$\rm ^a$         &  $S0_3$, \ion{H}{2}/LINER\\
     Center position$\rm ^a$    &  R.A. 15h06m29.5s\\
     ~~~~~(J2000)                &  Dec. +55d45m47.6s\\
     Distance (Mpc)$\rm ^a$     &  $11.21\pm0.80$ (1\farcs $\sim$ 54~pc)\\
     Inclination angle$\rm ^b$  &  $86.4^\circ$\\
     $D_{25}$ (arcmin)$\rm ^b$  &  $6.3\pm0.5$\\
     Total B magnitude$\rm ^b$ & $10.731\pm0.074$\\
     Total K magnitude$\rm ^a$ & $6.873\pm0.018$\\
     Total B-V color$\rm ^b$ & 0.85\\
     Foreground Galactic Extinction (E(B-V))$\rm ^a$ & 0.013\\
%Circular velocity ($km~s^{-1}$)$\rm ^b$ & $265\pm21$\\
     $M_{FIR}$$\rm ^b$  &  11.044\\
     Galactic foreground $N_H (10^{20}\rm~cm^{-2})^c$ & 1.46\\
     $M_{HI}$$\rm ^d$  &  $<1.2\times10^{8}\rm~M_\odot$\\
     $M_{H_2}$$\rm ^e$  &  $4.39\times10^{8}\rm~M_\odot$\\
     $M_{HII}$$\rm ^f$  &  $4\times10^{4}\rm~M_\odot$\\
     $M_{dust}$$\rm ^g$  &  $4.5\times10^{6}\rm~M_\odot$\\
     $L_{dust}$$\rm ^g$  &  $3\times10^{9}\rm~L_\odot$\\
     $f_{60}/f_{100}$$\rm ^h$ & 0.29\\
     SFR($\rm~M_\odot~yr^{-1}$)$\rm ^h$ & $<0.1$\\
     $q_{PAH}$$\rm ^g$  &  2\%\\
     $M_{dust}/M_H$$\rm ^g$  &  0.005\\

\enddata
\tablecomments{References. - a. NED; b. HYPERLEDA, $D_{25}$ is the
diameter at $I_B=25\rm~mag~arcsec^{-2}$, IRAS FIR magnitude
$M_{FIR}=-2.5~log(2.58~f_{60}+f_{100})+14.75$, where $f_{60}$, and
$f_{100}$ are the 60 and 100 $\mu$m fluxes, respectively, in Jy; c.
HEASARC web tools; d. Sage \& Welch 2006; e. Welch \& Sage 2003; f.
Plana et al. 1998; g. Draine et al. 2007. h. Kennicutt et al. 2003,
$f_{60}/f_{100}$ is the IRAS 60 $\mu$m to 100 $\mu$m flux
ratio.}\label{table:basicpara}
  \end{deluxetable}
  \vfill

\begin{deluxetable}{lrrrrrrrr}
  \tabletypesize{\footnotesize}
  \tablecaption{{\sl Chandra} Source List}
  \tablewidth{0pt}
  \tablehead{
  \colhead{Source} &
  \colhead{CXOU Name} &
  \colhead{$\delta_x$ ($''$)} &
  \colhead{CR $({\rm~cts~ks}^{-1})$} &
  \colhead{HR} &
  \colhead{HR1} &
%  \colhead{HR2} &
  \colhead{Flag} \\
  \noalign{\smallskip}
  \colhead{(1)} &
  \colhead{(2)} &
  \colhead{(3)} &
  \colhead{(4)} &
  \colhead{(5)} &
  \colhead{(6)} &
  \colhead{(7)}
%  \colhead{(8)} &
% \colhead{(9)}
  }
  \startdata
   1 &  J150552.44+554415.8 &  1.4 &$     1.10  \pm   0.32$& --& --& B, S \\
   2 &  J150555.82+554556.9 &  1.2 &$     1.57  \pm   0.39$& --& --& B, S \\
   3 &  J150559.07+554455.2 &  1.1 &$     1.28  \pm   0.35$& --& --& B, S \\
   4 &  J150600.69+554921.6 &  2.0 &$     0.42  \pm   0.18$& $ 1.00\pm0.04$ & --& H \\
   5 &  J150601.04+554513.0 &  0.9 &$     3.46  \pm   0.50$& $-0.03\pm0.18$ & $ 0.53\pm0.20$ & B, S, H \\
   6 &  J150603.95+554623.3 &  0.9 &$     1.28  \pm   0.31$& --& --& B, S \\
   7 &  J150605.82+554649.4 &  0.8 &$     4.70  \pm   0.67$& --& $-0.24\pm0.15$ & B, S, H \\
   8 &  J150608.65+554919.2 &  1.4 &$     0.40  \pm   0.17$& $ 1.00\pm0.02$ & --& H \\
   9 &  J150613.31+554359.4 &  0.7 &$     2.41  \pm   0.44$& --& --& B, S, H \\
  10 &  J150616.84+554715.2 &  0.5 &$     4.29  \pm   0.61$& $-0.26\pm0.18$ & $-0.06\pm0.16$ & B, S, H \\
  11 &  J150618.38+554547.7 &  0.7 &$     0.57  \pm   0.22$& --& --& B \\
  12 &  J150620.41+554530.3 &  0.6 &$     0.78  \pm   0.22$& --& --& B, H, S \\
  13 &  J150620.46+554622.7 &  0.4 &$     1.39  \pm   0.33$& --& --& B, S, H \\
  14 &  J150620.48+554542.2 &  0.4 &$     6.69  \pm   0.76$& $-0.27\pm0.14$ & $ 0.08\pm0.14$ & B, S, H \\
  15 &  J150620.66+554221.4 &  1.0 &$     0.46  \pm   0.17$& --& --& B, H \\
  16 &  J150620.76+554442.7 &  0.6 &$     0.37  \pm   0.20$& $-1.00\pm0.09$ & --& S \\
  17 &  J150621.30+554906.7 &  0.6 &$     5.27  \pm   0.74$& $-0.18\pm0.17$ & $-0.00\pm0.17$ & B, S, H \\
  18 &  J150621.36+554344.1 &  1.1 &$     0.44  \pm   0.22$& --& --& B \\
  19 &  J150622.50+554721.8 &  0.6 &$     1.02  \pm   0.31$& --& --& S, B \\
  20 &  J150623.05+554542.1 &  0.7 &$     0.82  \pm   0.28$& --& --& B, S \\
  21 &  J150624.53+554707.5 &  0.8 &$     0.57  \pm   0.20$& --& --& B, S, H \\
  22 &  J150625.18+554556.6 &  0.5 &$     0.77  \pm   0.23$& --& --& B, S \\
  23 &  J150627.77+554537.8 &  0.5 &$     0.89  \pm   0.27$& --& --& B, S, H \\
  24 &  J150628.11+554631.6 &  0.9 &$     0.87  \pm   0.27$& --& --& B, S \\
  25 &  J150628.43+554556.4 &  0.7 &$     0.92  \pm   0.29$& --& --& B \\
  26 &  J150628.73+554555.1 &  0.5 &$     2.10  \pm   0.41$& --& --& B, H, S \\
  27 &  J150628.75+554538.7 &  0.5 &$     1.83  \pm   0.37$& --& --& B, S, H \\
  28 &  J150629.14+554551.4 &  0.8 &$     0.56  \pm   0.21$& --& --& B \\
  29 &  J150629.15+554515.6 &  0.7 &$     0.54  \pm   0.20$& --& --& B, S \\
  30 &  J150629.51+554547.5 &  0.6 &$     0.61  \pm   0.20$& $ 1.00\pm0.10$ & --& H, B \\
  31 &  J150629.65+554552.7 &  1.2 &$     0.69  \pm   0.25$& --& --& B \\
  32 &  J150629.73+554309.0 &  0.4 &$     6.30  \pm   0.76$& $-0.59\pm0.14$ & $-0.02\pm0.13$ & S, B, H \\
  33 &  J150629.84+554533.2 &  0.6 &$     0.79  \pm   0.26$& --& --& B, S \\
  34 &  J150630.33+554648.8 &  0.6 &$     0.32  \pm   0.13$& $ 1.00\pm0.01$ & --& H, B \\
  35 &  J150630.43+554139.5 &  0.6 &$     8.65  \pm   1.10$& $-0.37\pm0.14$ & $ 0.03\pm0.15$ & B, S, H \\
  36 &  J150630.45+554515.4 &  0.7 &$     0.63  \pm   0.21$& --& --& B, H \\
  37 &  J150631.26+554458.1 &  0.5 &$     1.80  \pm   0.38$& --& --& B, S, H \\
  38 &  J150631.58+554251.3 &  0.6 &$     0.92  \pm   0.29$& --& --& B, S, H \\
  39 &  J150631.83+554232.3 &  0.7 &$     0.82  \pm   0.25$& --& --& B, S, H \\
  40 &  J150632.17+554220.7 &  0.6 &$     0.94  \pm   0.23$& $ 0.86\pm0.14$ & --& B, H \\
  41 &  J150632.70+554658.9 &  0.5 &$     1.96  \pm   0.35$& --& $ 0.81\pm0.18$ & B, S, H \\
  42 &  J150632.73+554525.6 &  0.6 &$     0.46  \pm   0.17$& --& --& B, H, S \\
  43 &  J150634.07+554536.6 &  0.7 &$     0.68  \pm   0.23$& --& --& B, H, S \\
  44 &  J150635.60+554445.6 &  1.7 &$     0.41  \pm   0.18$& --& --& B \\
  45 &  J150635.78+554531.4 &  0.6 &$     0.51  \pm   0.21$& $ 1.00\pm0.01$ & --& B, H \\
  46 &  J150637.04+554527.3 &  1.7 &$     0.33  \pm   0.18$& $ 1.00\pm0.02$ & --& B \\
  47 &  J150638.02+554842.2 &  0.5 &$     2.88  \pm   0.50$& --& --& B, S, H \\
  48 &  J150638.74+554505.6 &  0.6 &$     1.16  \pm   0.32$& --& --& B, S, H \\
  49 &  J150640.17+554612.5 &  1.4 &$     0.33  \pm   0.14$& --& --& H, B \\
  50 &  J150640.26+554514.0 &  0.9 &$     0.52  \pm   0.22$& --& --& B, S \\
  51 &  J150640.48+554806.6 &  0.4 &$     2.49  \pm   0.47$& --& --& S, B, H \\
  52 &  J150643.97+554522.9 &  0.7 &$     0.46  \pm   0.17$& --& --& B, H, S \\
  53 &  J150645.29+554845.3 &  0.6 &$     0.40  \pm   0.19$& $ 1.00\pm0.01$ & --& B \\
  54 &  J150652.51+554639.5 &  1.0 &$     0.45  \pm   0.19$& --& --& B, H \\
  55 &  J150653.12+554800.1 &  0.6 &$     0.70  \pm   0.24$& --& --& B, H \\
  56 &  J150653.79+554432.9 &  0.5 &$     1.32  \pm   0.55$& --& --& B, S, H \\
  57 &  J150655.64+554437.7 &  0.5 &$     0.56  \pm   0.19$& --& --& B, H \\
  58 &  J150656.64+554550.5 &  0.4 &$     1.99  \pm   0.38$& --& --& B, H, S \\
  59 &  J150701.15+554715.6 &  0.9 &$     1.16  \pm   0.67$& --& --& B, S \\
  60 &  J150703.66+554833.4 &  1.7 &$     0.93  \pm   0.54$& --& --& B \\
  61 &  J150703.70+554900.6 &  1.2 &$     0.59  \pm   0.23$& --& --& B, H \\
  62 &  J150704.60+554448.9 &  0.7 &$     2.93  \pm   1.06$& --& --& S, B \\
  63 &  J150707.34+554435.4 &  1.0 &$     0.44  \pm   0.18$& $ 1.00\pm0.01$ & --& H, B \\
  64 &  J150707.50+554527.1 &  1.0 &$     1.53  \pm   0.73$& --& --& B \\
  65 &  J150708.00+554613.9 &  0.6 &$    12.63  \pm   2.11$& $-0.30\pm0.15$ & $-0.42\pm0.13$ & B, S, H \\
  66 &  J150708.48+554545.8 &  0.7 &$     1.33  \pm   0.32$& --& --& B, H \\
  67 &  J150708.78+554144.7 &  1.3 &$     2.34  \pm   0.69$& $ 1.00\pm0.00$ & --& B, H \\
  68 &  J150715.72+554955.2 &  2.1 &$     1.03  \pm   0.40$& --& --& B, S \\
  69 &  J150719.26+554449.3 &  1.0 &$     6.12  \pm   1.16$& $-0.13\pm0.17$ & --& B, S, H \\
  70 &  J150721.08+554559.0 &  1.0 &$    45.48  \pm   3.75$& $-0.31\pm0.07$ & $-0.25\pm0.08$ & B, S, H \\
  71 &  J150736.05+554605.6 &  2.3 &$     1.85  \pm   0.81$& --& --& B, S \\
\enddata
\tablecomments{The definition of the bands:
0.3--0.7 (S1), 0.7--1.5 (S2), 1.5--3 (H1), and 3--7~keV (H2). % for ACIS-S
%0.5--1 (S1), 1--2 (S2), 2--4 (H1), and 4--8~keV  (H2). % for ACIS-I
%1--2.5 (S1), 2.5--4 (S2), 4--6 (H1), and 6--9~keV  (H2). % for ACIS-I_low
In addition, S=S1+S2, H=H1+H2, and B=S+H.
 Column (1): Generic source number. (2):
{\sl Chandra} X-ray Observatory (unregistered) source name,
following the {\sl Chandra} naming convention and the IAU
Recommendation for Nomenclature (e.g.,
http://cdsweb.u-strasbg.fr/iau-spec.html). (3): Position uncertainty
(1$\sigma$) calculated from the maximum likelihood centroiding and
an approximate off-axis angle ($r$) dependent systematic error
$0\farcs2+1\farcs4(r/8^\prime)^2$ (an approximation to Fig.~4 of
Feigelson et al.~2002), which are added in quadrature.
%\bibitem[Feigelson et al. (2002)]{fei02} Feigelson, E., et al. 2002, ApJ, 574, 258
(4): On-axis source broad-band count rate --- the sum of the
exposure-corrected count rates in the four
bands. (5-6): The hardness ratios defined as
%${\rm HR}=({\rm H1-S})/({\rm H1+S})$, and ${\rm HR1}=({\rm H2-H1})/{\rm H}$, %dhrch=2
${\rm HR}=({\rm H-S2})/({\rm H+S2})$, and ${\rm HR1}=({\rm S2-S1})/{\rm S}$, %dhrch=0
listed only for values with uncertainties less than 0.2.
(7): The label ``B'', ``S'', or ``H'' mark the band in
which a source is detected with the most accurate position that is adopted in
Column (2).
%The label ``v'' denotes that a source is a variable.
}\label{table:acis_source_list}
  \end{deluxetable}
  \vfill

\begin{deluxetable}{lrrrrrrrr}
  \tabletypesize{\footnotesize}
  \tablecaption{Properties of Individual Dust Features}
  \tablewidth{0pt}
  \tablehead{
  \colhead{Feature} & {Feature} & {$|z|$} & {$a_B$} & {$a_V$} & {$N_H$} & {$n$} & {$M$} & {$\Omega$}\\
No. & ID & (pc) & (mag) & (mag) & ($10^{20}\rm~cm^{-2}$) & ($\rm cm^{-3}$) & ($10^4\rm~M_\odot$) & ($10^{51}\rm~ergs$)\\
(1) & (2) & (3) & (4) & (5) & (6) & (7) & (8) & (9)\\
    }
  \startdata
    01 & $D-012+002$  & 110 & 0.55 & 0.45 & 8.0  & 3.3 & 5  & 3  \\
    02 & $D-008+003$  & 150 & 0.41 & 0.32 & 5.7  & 2.3 & 10 & 10 \\
    03 & $D-004+002$  & 100 & 0.78 & 0.56 & 10.1 & 6.2 & 5  & 2  \\
    04 & $D+000+003$  & 140 & 0.31 & 0.22 & 3.9  & 4.9 & 2  & 2  \\
    05 & $D+001+005$  & 270 & 0.10 & 0.07 & 1.2  & 0.9 & 1  & 3  \\
    06 & $D+002+002$  & 120 & 0.47 & 0.37 & 6.7  & 8.3 & 2  & 1  \\
    07 & $D+008+004$  & 190 & 0.17 & 0.11 & 2.0  & 1.0 & 2  & 3  \\
    08 & $D+004-001$  & 70  & 0.39 & 0.29 & 5.1  & 3.2 & 1  & 0.3\\
    09 & $D+001-001$  & 80  & 0.16 & 0.11 & 2.0  & 2.5 & 0.5& 0.2\\
    10 & $D-001-001$  & 70  & 0.43 & 0.31 & 5.5  & 4.6 & 1  & 0.3\\
    11 & $D-003-002$  & 90  & 0.36 & 0.22 & 4.0  & 3.8 & 2  & 0.6\\
    12 & $D-008-001$  & 80  & 0.36 & 0.29 & 5.3  & 2.8 & 3  & 0.8\\
    13 & $D-010-001$  & 70  & 0.40 & 0.32 & 5.7  & 2.8 & 4  & 0.9\\
    14 & $D-013-001$  & 60  & 0.42 & 0.31 & 5.6  & 7.0 & 2  & 0.3\\
    15 & $D-016-002$  & 90  & 0.41 & 0.32 & 5.8  & 3.6 & 4  & 2  \\
    16 & $D+008+014$  & 780 & 0.14 & 0.11 & 1.9  & 1.2 & 10 & 17 \\

\enddata
\tablecomments{(1) Dust feature numbers as marked in
Fig.~\ref{fig:dustfeature}. (2) Feature identification in form of
$D\pm XXX\pm ZZZ$, where XXX is the distance in arcseconds from the
galaxy's optical center along the major axis, while ZZZ is the
height in arcseconds from the mid-plane of the galaxy. (3) Physical
height of the dust features from the mid-plane. (4) and (5) Apparent
extinction in the B and V band. (6) Hydrogen column density. (7)
Number density. (8) Gas mass. (9) Gravitational
energy.}\label{table:dustfeature}
  \end{deluxetable}
  \vfill

\end{document}